\documentclass[aps,prb,preprint,groupedaddress,superscriptaddress, showpacs, ,showkeys]{revtex4-1}
\usepackage{amssymb}
\usepackage{amsmath}
\usepackage{amsfonts}
\usepackage{amstext}
\usepackage{amsgen}
\usepackage{amsbsy}
\usepackage{amsopn}
\usepackage{keyval}
\usepackage{trig}
\usepackage{graphicx}
\usepackage{graphics}
\usepackage{bm}
\usepackage{natbib}
\usepackage{textcase}

\begin{document}

\title{The influence of the elementary charge on the canonical quantization of $LC$ -circuits}

\author{E. Papp}
\email[E. Papp]{, Email:  erpapp@yahoo.com}
\affiliation{Physics Department, West University of Timisoara, RO-300223, Timisoara, Romania}

\author{D. Racolta}
\affiliation{Faculty of Science, North University of Baia Mare, RO-430122, Baia Mare, Romania}
\author{L. Aur}
\email[L. Aur]{, Email:  liviuaur@gmail.com}
\affiliation{Physics Department, West University of Timisoara, RO-300223, Timisoara, Romania}
\author{Z. Szakacs}
\affiliation{Faculty of Science, North University of Baia Mare, RO-430122, Baia Mare, Romania}

\date{\today}

\begin{abstract}

In this paper one deals with the quantization of mesoscopic
$LC$-circuits under the influence of an external time dependent voltage. The
canonically conjugated variables, such as given by the electric charge and the
magnetic flux, get established by resorting to the hamiltonian equations of
motion provided by both Faraday and Kirchhoff laws . This time the
discretization of the electric charge is accounted for, so that magnetic flux
operators one looks for should proceed in terms of discrete derivatives.
However, the flux operators one deals witg are not Hermitian, which means that
subsequent symmetrizations are in order. The eigenvalues characterizing such
operators cab be readily established in terms of twisted boundary conditions.
Besides the discrete Schr\"{o}dinger equation with nearest-neighbor hoppings,
a nontrivial next nearest neighbor generalization has also been established.
Such issues open the way to the derivation of persistent currents in terms of
effective $k$-dependent Hamiltonians. Handling the time dependent voltage
within the nearest neighbor description leadsto the derivation of dynamic
localization effects in $L$-ring configurations, such as discussed before by
Dunlap and Kenkre The onset of the magnetic flux quantum has also been
discussed in some more detail.
\end{abstract}



\maketitle

\section{INTRODUCTION}

The increasing miniaturization of electric mesoscopic devices attains a stage
in which the length scale of the $LC$system becomes smaller than the so called
phase coherence length, which stands for the length scale characterizing
quantum interference devices. This means that the quantization ofthe
$LC$-circuit is in order [1]. We shall then proceed by applying the canonical
quantization, now by selecting canonically conjugated variables one deals with
in an appropriate manner, such as given by the electric charge $Q$ and the
rescaled magnetic $\Phi/c$ [2]. For this purpose we have to remind that the
energy terms concerning the capacitance $C $ and the inductance $L$ are given
by $Q^{2}/2C$ and $\Phi^{2}/2Lc^{2}$, respectively. Accordingly, one deals
with the total Hamiltonian%

\begin{equation}
H(Q,\Phi/c)=\frac{\Phi^{2}}{2Lc^{2}}+\frac{Q^{2}}{2C}-QV_{s}(t)\tag{1}%
\end{equation}
in which the influence of an external time dependent voltage has also been
included. Accordingly, the Hamiltonian equations of motion are given by
$\Phi=ILc$ and%

\begin{equation}
U=V_{s}(t)-\frac{d}{dt}\left(  \frac{\Phi}{c}\right) \tag{2}%
\end{equation}
where $I=dQ/dt$ and $U=Q/C$. This supports the selection of canonically
conjugated variables just done above. Just remark that (2) encompasses both
Faraday's law of induction and the second law of Kirchhoff.

So, one gets faced with the canonical commutation relation%

\begin{equation}
\left[  Q_{op},\Phi_{op}\right]  =i\hbar c,\tag{3}%
\end{equation}
where by now $Q_{op}$and $\Phi_{op}$ denote quantum-mechanical observables
concerningthe electric charge and the magnetic flux, respectively. In this
context Eq.(3) is responsible for the quantization of the $LC$-circuit.

Next let us look for a solution of Eq.(3), now by accounting for the existence
of an elementary charge, say $q_{e}=e=1.60217733\times10^{-19}C$, with the
understanding that the charge of the electron is $-e$. We then have to say
that the charge $Q$ is a discrete one if $Q=nq_{e}$, where $n$ denotes an
integer. Our next task is to perform the quantization of the $LC$-circuit
referred to above in terms of the right-hand and left-hand discrete
derivatives $\Delta$ and $\nabla$ for which
\begin{equation}
\Delta f(n)=f(n+1)-f(n)\quad and\quad\nabla f(n)=f(n)-f(n-1)\tag{4}%
\end{equation}
in which case $\Delta^{+}=-\nabla$ and $\Delta\nabla=\Delta-\nabla$. We have
to say that the $n$-integer is provided by the eigenvalue equation [3]%

\begin{equation}
Q_{op}\mid n>=Q\mid n>\tag{5}%
\end{equation}
where $Q=nq_{e}$. Disregarding for the moment Eq. (5), leads us to say that
the canonical commutation relation (3) exhibits the solution%

\begin{equation}
\Phi_{op}=-i\hbar c\frac{\partial}{\partial Q}=-i\frac{\hbar c}{q_{e}}%
\frac{\partial}{\partial n}\tag{6}%
\end{equation}
in so far as $Q$ is a continuous variable. We have to remark that the quotient
$\delta\Phi=\hbar c/q_{e}$ $=h/2q_{e}(c/2\pi)$ in(6) serve as a candidate for
the magnetic flux quantum, but further clarifications remain desirable.
Indeed, the magnetic flux quantum, say $\phi_{0}$, is registered in data
tables of fundamental constants as $\phi_{0}=$ $h/2e=2.06783372\cdot
10^{-15}Wb$, in which case $\delta\Phi/\phi_{0}=c/2\pi$. Moreover, the
magnetic flux quantum, in CGS-units, such as used in connection with the
Aharonov-Bohm effect reads $\phi_{A}=hc/q_{e}$. Of course, $\phi_{0}$ differs
both from $\delta\Phi$ and $\phi_{A}$ ,which indicates that a\ safe
theoretical derivation of the magnetic flux quantum is in order.

\section{ MAGNETIC\ FLUX-STRUCTURES AND\ MAGNETIC\ ENERGIES}

The derivation of the discrete counterpart of (6) requires a little bit more
attention. Indeed, in this case on gets faced with two non-Hermitian
realizations of the magnetic flux -operator, say [4]%

\begin{equation}
\Phi_{op}^{(1)}=\frac{-i\hbar c}{q_{e}}\nabla\tag{7}%
\end{equation}
and%

\begin{equation}
\Phi_{op}^{(2)}=\frac{-i\hbar c}{q_{e}}\Delta=\left(  \Phi_{op}^{(1)}\right)
^{+}.\tag{8}%
\end{equation}
Accordingly, the symmetrized Hermitian magnetic flux operator is given by%

\begin{equation}
\Phi_{sym}=\frac{-i\hbar c}{2q_{e}}\left(  \Delta+\nabla\right)  .\tag{9}%
\end{equation}
Then the eigenvalues of such operators should provide realizations of a
quantized magnetic flux. In addition, the Hermitian flux operator $\Phi_{sym}$
yields a symmetrized magnetic energy via
\begin{equation}
\mathit{H}_{sym}=\mathit{H}_{NNN}=\frac{1}{2Lc^{2}}\Phi_{sym}^{2}=-\frac
{\hbar^{2}}{8Lq_{e}^{2}}\left(  \Delta+\nabla\right)  ^{2},\tag{10}%
\end{equation}
which is responsible for next nearest-neighbor (NNN) hoppings.

In addition, we have to look for an Hermitian magnetic energy operator like%

\begin{equation}
\mathit{H}_{m}=\mathit{H}_{NN}=\frac{1}{2Lc^{2}}\left(  \Phi_{op}%
^{(1)}\right)  ^{+}\Phi_{op}^{(1)}=-\frac{\hbar^{2}}{2Lq_{e}^{2}}\left(
\Delta-\nabla\right)  ,\tag{11}%
\end{equation}
which is provided by a different kind of symmetrization and wich is
responsible for nearest-neighbor (NN) hoppings. So one gets faced with two
competing Hermitian realizations, i.e. with $\mathit{H}_{NN}$ and
$\mathit{H}_{NNN}$ , proceeding solely in a direct connection with the
underlying symmetrization. In other words we have to account for an actual
sensitivity with respect to the symmetrization path, which looks interesting
from a theoretical point of view. We have also to mention that under the
discretization of the electric charge the canonical commutation relation (3)
gets modified as%

\begin{equation}
\left[  Q_{op},\Phi_{sym}\right]  =-i\hbar c\left(  1-\frac{q_{e}^{2}L}%
{\hbar^{2}}\mathit{H}_{m}\right) \tag{12}%
\end{equation}
which can be viewed as a deformed algebraic structure.Note that such
structures have received much attention during the last $2-3$ decades of the
former century [5].

The energy dispersion laws characterizing $\mathit{H}_{NN}$ and $\mathit{H}%
_{NNN}$, say $E_{NN}(k)$ and $E_{NNN}(k)$, can also be readily established.
One obtains%

\begin{equation}
E_{NN}(k)=-\frac{\hbar^{2}}{Lq_{e}^{2}}\left(  \cos k-1\right) \tag{13}%
\end{equation}
and%

\begin{equation}
E_{NNN}(k)=\frac{\hbar^{2}}{2Lq_{e}^{2}}\sin^{2}k\tag{14}%
\end{equation}
which are both even functions of the $k$-parameter, as one might expect.

\section{REVISITING\ THE\ DISCRETE NN- and
NNN-\ SCHR\"{O}DINGER\ EQUATIONS\ OF\ THE\ $LC$-CIRCUIT}

Next let us discuss the Schr\"{o}dinger equation of the $LC$-circuit by
resorting to the magnetic energy $\mathit{H}_{m}$ such as written down above
in (11). This equation exhibits the general form%

\begin{equation}
\mathit{H}_{m}\mid\Psi>=i\hbar\frac{\partial}{\partial t}\mid\Psi>\tag{15}%
\end{equation}
which will be handled in terms of the wavefunction%

\begin{equation}
\varphi_{m}(t)=<m\mid\Psi>.\tag{16}%
\end{equation}
where $m$ is an integer. This leads to the second order discrete equation with NN-hoppings%

\begin{equation}
-\frac{\hbar^{2}}{2Lq_{e}^{2}}\left(  \varphi_{m+1}+\varphi_{m-1}\right)
+\left(  \frac{q_{e}^{2}}{2C}m^{2}-mq_{e}V_{s}\left(  t\right)  +\frac
{\hbar^{2}}{Lq_{e}^{2}}\right)  \varphi_{m}=i\hbar\frac{\partial}{\partial
t}\varphi_{m},\tag{17}%
\end{equation}
which looks like a perturbed harmonic oscillator on the discrete space. One
realizes that the constant $\hbar^{2}/Lq_{e}^{2}$-term in (17) can be gauged
away. In addition,(17) shows that the NN overlap integral is given by
$J_{NN}=-\hbar^{2}/2Lq_{e}^{2}$. Acounting for NNN-hoppings,one realizes that
(17) gets replaced by%

\begin{equation}
-\frac{\hbar^{2}}{8Lq_{e}^{2}}\left(  \varphi_{m+2}+\varphi_{m-2}\right)
+\left(  \frac{q_{e}^{2}}{2C}m^{2}-mq_{e}V_{s}\left(  t\right)  +\frac
{\hbar^{2}}{4Lq_{e}^{2}}\right)  \varphi_{m}=i\hbar\frac{\partial}{\partial
t}\varphi_{m},\tag{18}%
\end{equation}
by virtue of (10). One sees that $J_{NN}$ gets modified as $J_{NNN}=-\hbar
^{2}/8Lq_{e}^{2}$ . It is understood that such issues have to be viewed as
first principles results.

Neglecting the term quadratic in the discrete coordinate $m$, it can be easily
verified that the Fourier transform%

\begin{equation}
\varphi_{m}=\frac{1}{2\pi}\int dkC_{k}(t)\exp(imk).\tag{19}%
\end{equation}
leads to the conversion of (17) into%

\begin{equation}
\left(  -\frac{\hbar^{2}}{Lq_{e}^{2}}\cos k-iq_{e}V_{s}\left(  t\right)
\frac{\partial}{\partial k}\right)  C_{k}(t)=i\hbar\frac{\partial}{\partial
t}C_{k}(t),\tag{20}%
\end{equation}
which works in the $k$-representation. A similar transformation can also be
done for (18). However,it should be remarked that the Hamiltonian
characerizing (20) is not Hermitian for realistic systems for which
$V_{s}\left(  t\right)  \neq0$, unless one considers a time lattice in which
$V_{s}\left(  t\right)  =0$. Now we have to say that more detailed
transformations providing equivalent Hamiltonians which are both $k$- and
flux-dependent have been discussed [6]. We shall then apply this latter
approach to the study of the magnetic flux quantum, now by accounting in an
explicit manner for the $k$-parity of the Hamiltonian. Proceeding in this
manner opens the way to a proper derivation of the magnetic flux quantum
$\phi_{0}$ one looks for. For this purpose one resorts to the shifted
Fourier-transform
\begin{equation}
u(k,t)=%
%TCIMACRO{\dsum \limits_{n=-\infty}^{n=\infty}}%
%BeginExpansion
{\displaystyle\sum\limits_{n=-\infty}^{n=\infty}}
%EndExpansion
C_{n}(t,k)\exp in\left(  k+\frac{q_{e}\Phi_{e}}{\hbar c}\right) \tag{21}%
\end{equation}
in which case (17) gets converted into
\begin{equation}
H_{eff}(k,\Phi)C_{n}(t,k)=i\hbar\frac{\partial}{\partial t}C_{n}(t,k),\tag{22}%
\end{equation}
where%

\begin{equation}
H_{eff}(k,\Phi)=-\frac{q_{e}^{2}}{2C}\frac{\partial^{2}}{\partial k^{2}}%
+\frac{\hbar^{2}}{q_{e}^{2}L}\left(  1-\cos\left(  k+\frac{q_{e}\Phi_{e}%
}{\hbar c}\right)  \right)  ,\tag{23}%
\end{equation}
denotes the effective $k$-dependent Hamiltonian and where $\Phi_{e}$ stands
for the external flux. The persistent current is then given by
\begin{equation}
I_{k}(t)=-c\frac{\partial H_{eff}}{\partial\Phi_{e}}=-\frac{\hbar}{q_{e}L}%
\sin\left(  k+\frac{q_{e}\Phi_{e}}{\hbar c}\right)  .\tag{24}%
\end{equation}
Assuming that the effective Hamiltonian displayed above is an even function of
$k$, yields the flux quantization rule one looks for as%
\begin{equation}
\frac{q_{e}\Phi_{e}}{hc}=\frac{n}{2},\tag{25}%
\end{equation}
where $n$ is an integer. This shows that the magnetic flux quantum is given
precisely by $\Phi_{0}=hc/2q_{e}$, with the understanding that the $c$-factor
is reminiscent to CGS-units.

\section{APPLYING\ MAGNETIC\ FLUX\ OPERATORS\ IN\ TERMS\ OF\ TWISTED\ BOUNDARY\ CONDITIONS
\ \ \ \ \ \ }

The motion of an electron on a 1-D ring threaded by an external magnetic flux
$\Phi_{e}$ is characterized by the twisted boundary condition [7]%

\begin{equation}
\varphi(x+L)=\exp2\pi i\left(  \beta+n\right)  \varphi(x),\tag{26}%
\end{equation}
in which $n=0,\pm1,\pm2,...$and $\beta=\Phi_{e}/\Phi_{0}$.The electron
coordinate along the ring and the ring circumference are denoted by $x$ and
$L$, respectively. One has%

\begin{equation}
\Delta\varphi(x)=\varphi(x+L)-\varphi(x)=2i\exp i\pi\left(  \beta+n\right)
\sin\pi\left(  \beta+n\right)  \varphi(x),\tag{27}%
\end{equation}
so that%

\begin{equation}
\Phi_{op}^{(2)}\varphi(x)=\frac{-i\hbar c}{q_{e}}\Delta\varphi(x)=\frac{2\hbar
c}{q_{e}}\exp\left(  i\pi(\beta+n\right)  )\sin\left(  \pi(\beta+n\right)
)\varphi(x).\tag{28}%
\end{equation}
Proceeding in a similar manner one finds%

\begin{equation}
\Phi_{sym}\varphi(x)=\frac{\hbar c}{q_{e}}\sin\left(  2\pi(\beta+n\right)
)\varphi(x).\tag{29}%
\end{equation}
We can then say that the eigenvalues of the flux operators one deals with can
be established actually in terms of Eq.(26), which serves for a deeper
understanding. However, further clarifications remain desirable.

\section{CONCLUSIONS}

In this paper the canonical quantization of the mesoscopic$LC$-circuit has
been discussed under specific conditions concerning the discrete spectrum of
the electric charge as well as the influence of an external time dependent
voltage. This amounts to formulate the magnetic flux operator in terms of
right- and left-hand discrete derivatives, as shown by (7) and (8). Such
operators are not hermitian owing to the very incorporation of discrete
derivatives, but a hermitian realization can be readily established by
resorting to a subsequent symmetrization, such as displayed in (9). The
eigenvalues of such operators have been established by resorting to twisted
boundary conditions. A deformed canonical commutation relation between
electric charge and magnetic flux such as given by (12) has also been derived.
This is a nontrivial result which deserves further attention from the
theoretical point of view. The derivation of the discrete Schr\"{o}dinger
equations of the $LC$-circuit, say (17) and (18), by accounting for the
magnetic energy operators $H_{m}=H_{NN}$ and $H_{sym}=H_{NNN}$, should also be
mentioned. Moreover, these equations exhibit the attributes of
first-principles descriptions as they provide both the hopping configurations
as well as the corresponding overlap integrals. And last but not at least a
safe derivation of the magnetic flux qantum has also been established as shown
by Eq.(25). To this aim one should deal with effective Hamiltonians which are
even functions of the wavenumber $k$.

\section*{ACKNOWLEDGMENTS}

We\ remain\ deeply\ indebted\ to\ the\ late\ Professor Micu Codrutza\ for\ her\ unvaluable\ supports\ and\ encouragements.

\end{document}